\documentclass[12pt,preprint]{aastex}

\newcommand{\cxo}{{\sl Chandra}}

\newcommand{\xmm}{{\sl XMM-Newton}}
\newcommand{\hst}{{\sl Hubble}}
\newcommand{\msun}{M$_{\odot}$}
\newcommand{\ergl}{ergs~s$^{-1}$}

\newcommand{\ergcms}{ergs~cm$^{-2}$~s$^{-1}$}

\newcommand{\ros}{{\sl ROSAT}}

\newcommand{\rxh}{1RXH~J132519.8$-$430312}
\newcommand{\abe}{A~0538$-$66}

\begin{document}

\title{On the nature of the ultraluminous X-ray transient in Cen~A (NGC~5128)}

\author{
Kajal~K.~Ghosh\altaffilmark{1},
Mark H. Finger\altaffilmark{1},
Douglas~A.~Swartz\altaffilmark{1},
Allyn~F.~Tennant\altaffilmark{2}, and
Kinwah~Wu\altaffilmark{3} 
}
\altaffiltext{1}{Universities Space Research Association,
NASA Marshall Space Flight Center, XD12, Huntsville, AL, USA}
\altaffiltext{2}{Space Science Department,
NASA Marshall Space Flight Center, XD12, Huntsville, AL, USA}
\altaffiltext{3}{MSSL, University College London, Holmbury St. Mary, Surrey,
RH5 6NT, UK}%; and School of Physics, University of Sydney, NSW 2006, Australia}

\begin{abstract}

We combine   9 \ros, 9 \cxo, and 2 \xmm~ observations of the Cen~A galaxy  
 to obtain the X-ray light curve of \rxh ~ 
 (=CXOU J132519.9$-$430317) spanning 1990 to 2003. 
The source reached a peak 0.1--2.4~keV flux $F_{\rm X}$$>$$10^{-12}$~\ergcms\ 
 during a 10~day span in 1995 July.
The inferred peak isotropic luminosity of the source therefore
 exceeded $3\times 10^{39}$~ergs~s$^{-1}$,  
 which places the source in the class of ultra-luminous X-ray sources.
Coherent pulsations at 13.264 Hz 
 are detected during a second bright episode 
 ($F_{\rm X}$$>$$3\times10^{-13}$~\ergcms) in 1999 December. 
The source is detected and varies significantly 
 within three additional observations but is below the detection threshold in 7 observations. 
The X-ray spectrum in 1999 December is best described as a cut-off power law or a 
 disk-blackbody (multi-colored disk).
We also detect an optical source, $m_{\rm F555W}$$\sim$24.1 mag, within the \cxo\ error 
circle of \rxh\ in \hst\ images taken 195~days before the nearest X-ray observation. The optical brightness of this source is consistent with a
late O or early B star at the distance of Cen~A.
 If the optical source is the counterpart, then the X-ray and optical behavior of \rxh\ are similar to 
the transient Be/X-ray pulsar \abe.

\end{abstract}
%%%%%%%%%%%%%%%%%%%%%%%%%%%%%%%%%%%%%%%%%%%%%%%%%%%%%%%%%%%%%%%%%%%%%%%%%%%%%
\section{Introduction}
     
 X-ray transients in our Galaxy and the Magellanic clouds have been
observed since the beginning of X-ray astronomy. In systems with
low-mass stellar companions, accretion is generally through an accretion
disk which is fed by Roche lobe overflow of the companion. The transient
outbursts are believed to be due either to an instability in the 
accretion
disk similar to the thermal-viscous instability which causes dwarf nova 
outbursts
in catacylsmic variables (van Paradijs 1996), or to an instability in the 
companion's mass loss
due to heating of its atmosphere by radiation from the X-ray
source (Hameurty et al. 1986). 
  Many of these soft X-ray transient (SXT) outbursts are extremely bright,  
with one of the most luminous that from SAX J1819.3-2525 (V4641 Sgr) 
which reached an X-ray luminosity of $\sim 6\times 10^{39}$~\ergl, or $\sim$7 
times the Eddington luminosity for this 7~\msun\ black hole 
(McClintock \& Remillard 2003).

High-mass X-ray binary transient sources, on the other hand, 
are normally Be/X-ray binaries
in which a high magnetic field neutron star accretes from the circumstellar
outflow of the Be star (Verbunt \& van~den~Heuvel 1995). 
There are two forms of outburst: type~I which
occur near periastron passage in wide eccentric orbit systems,
tend to recur
for a number of orbits in sequence, and reach luminosities of 
$10^{35-37}$~\ergl; and type~II which occur singly, may last for 
multiple orbits, and often reach the Eddington luminosity (Negueruela 1998). 
The most luminous observed outburst was from A~0538$-$66 (1A~0535$-$668) which 
reach a peak luminosity of $L_x = 8.5\times 10^{38}$ ~\ergl~in the 2-17 keV band (White
\& Carpenter 1978). 

Identifying luminous transients in other galaxies is rare because
of source confusion within the host galaxy, the infrequency of X-ray observations of 
individual galaxies,
and the relative faintness of even the brightest non-nuclear point sources in nearby 
galaxies.
The transient \rxh\ discovered by Steinle, Dennerl \& Englhauser (2000) in \ros\ images of Cen~A (NGC~5128)  is an exception: 
\rxh\ is in a relatively isolated region of Cen~A away from the nucleus and the 
 conspicuous dust lane, there have been numerous deep X-ray observations of Cen~A, 
 and Cen~A is only 
4~Mpc distant (Ferrarese et al.\ 2000; Hui et al.\ 1993).
Previous analysis of the \ros\ observations of this source were reported by Steinle et al. (2000)
 and of a subset of the \cxo\ observations by Kraft et al. (2001).

In this paper we assemble the long-term X-ray light curve of \rxh\ spanning 1990 to 2003 (\S~2).
Our analysis accounts for a weak nearby source visible in \cxo\ images (see also Kraft et al. 2001) that is unresolved in 
 \ros\ and \xmm\ observations. 
We use the high time resolution \ros\ and \cxo/High Resolution Camera (HRC) observations 
 to search for coherent pulsations and describe our procedure and results in \S~3. 
Analysis of available \hst\ images is presented in \S~4. 
We summarize our analysis in \S~5 and show that the data favor the 
 interpretation of \rxh\ as an example of a Be/X-ray binary in Cen~A like \abe.

\section{The 1990-2003 X-ray light curve of \rxh}

The chronology of X-ray observations of \rxh\ is presented in Table~1. 
We estimated the fluxes listed in Table~1 
 by scaling from the only observation made with moderate spectral resolution while the source was bright;
 the \cxo/ACIS-I observation of 1999 Dec 5 (Table~1).
Figure~1 displays the 1999 Dec 5 spectrum and the best-fit model.
Table~2 lists all the spectral models we applied, fit parameter values, and fit statistics. 
Except the {\tt powerlaw} model, all trial models produced acceptable fits.
The cutoffpl ({\tt powerlaw} $*$ {\tt highecut}) or {\tt diskbb} models produced the lowest $\chi^2$ statistic.
For all the other observations,
we used the bremsstrahlung spectral fit parameters as inputs to PIMMS (Mukai 1993) to 
convert observed count rates to estimated fluxes in the 0.1--2.4~keV band.

%-----------------------------Figure Start--------------------------------

\begin{center}
\includegraphics[angle=-90,width=\columnwidth]{Fig1.ps}
\figcaption{The \cxo/ACIS-I spectrum of \rxh\ obtained on 1999 December 05 is shown along with
  the best-fitting abosrbed cut-off power law (powerlaw~\*~highecut) model and fit residuals. Parameter values and fit 
  statistics for this and other trial models are listed in Table~2.}
\end{center}

%-----------------------------Figure End---------------------------------- 

Figure~2 displays a portion of two \cxo/ACIS images of the Cen~A field that illustrate the 
 transient nature of \rxh\ and the presence of a weak nearby source. 
The weak source is CXOU~J132520.1-430310 (Kraft et al. 2001).
We estimated the flux from this source to be 
 $(8.6\pm1.1)\times 10^{-15}$~\ergcms\
 in the 2000 May 17 \cxo/ACIS-I observation.

Our re-analysis of \ros\ data ({\sl cf.} Steinle, Dennerl \& Englhauser 2000) took advantage of the
 precise \cxo\ location of \rxh\ relative to the nearby bright object H13 (  Fig. 2; see also Turner et al. 1997). see
Using this offset, we then extracted the counts from a circular region of 30\arcsec\ radius 
   centered at the transient position 
   in the \ros\ images and subtracted a background   
   determined  from a circular annulus extending from 30\arcsec\ to 60\arcsec\ radii, 
   also centered at the source position.
Our resulting flux values are systematically lower than  
   those obtained by Steinle, Dennerl \& Englhauser (2000) primarily because they
   assumed a power law with $\Gamma = 1.5$ and intervening absorption column of 
   $N_{\rm H} = 8 \times 10^{20}$~cm$^{-2}$, in their derivation in contrast to our bremsstrahlung 
   model (Table~2). 
A marginal detection in  the 1994 August 10 observation  was attributed to the weak nearby source seen by Chandra.

%-----------------------------Figure Start--------------------------------
\begin{center}
\includegraphics[angle=-90,width=\columnwidth]{Fig2a.ps}
\includegraphics[angle=-90,width=\columnwidth]{Fig2b.ps}
\figcaption{6.\arcmin4$\times$4.\arcmin0 portions of \cxo/ACIS-I observations of Cen~A obtained on 
1999 December 5 ({\sl top}) and 2000 May 17 ({\sl bottom}).
North is up and east is to the left.
The 20\arcsec$\times$20\arcsec\ box centered on \rxh\ represents the HST field displayed in Figure 8. 
There is a nearby weak source, CXOU~J132520.1-430310, 
 visible most clearly in the latter image within the box. 
This source is blended with \rxh\ in lower-resolution \ros\ and \xmm\ observations.
Note the readout streak from the bright Cen~A nucleus indicating  the  spacecraft roll angle change. The bright object at the lower right hand corner of each image is H13 (Turner et al. 1997), 
which was used to precisely determine the location of \rxh\ in \ros\ and \xmm\ observations. }
\end{center} 
%-----------------------------Figure End---------------------------------- 

For the \cxo\ observations,
   we extracted counts from an ellipse of half width 9\arcsec\ and half height 6\arcsec\ with a
   position angle dependent on the spacecraft roll at the time of observation. 
Corresponding background counts were extracted from the same size ellipse centered at three different 
 positions around the transient and then averaged.
For the ACIS/HETG observations, only the zeroth-order images were analyzed.

For the \xmm\ observations, we used \xmm\ Science Analysis System - GUI version 1.52.7 
to process the EPIC MOS and pn images.  The position of the transient was determined with respect to the precise location of H13 (Turner et al. 1997). Counts from the source and the background  were extracted from a circle of radius 10\arcsec\ and a circular annulus of 10\arcsec\ and 20\arcsec\ radii, respectively, at the position of \rxh\, in the 0.2 -- 10 keV band. 
   
We show in Figure~3 the combined 0.1--2.4 keV light curve of \rxh.
The source was bright  during a 10 day period in 1995 July and in two observations seperated by 16 days in 1999 December
(see also Table~1).
Three months prior to the 1999 December observations, the source was detected but at less than 
10\% of the flux it attained in 1999 December. 
The source faded to below the detection
 threshold in all subsequent observations except during the 2001 May 21 ACIS/HETG observation where
 the source was detected at a flux of only $3.5\times10^{-15}$~\ergcms. 
 For non-detections, we show 2$\sigma$ upper limits in Table~1 and Figure 3.
  
%-----------------------------Figure Start--------------------------------
\begin{center}
\includegraphics[angle=-90,width=\columnwidth]{Fig3.ps}
\figcaption{The  0.1$-$2.4 keV  light curve of  \rxh\
    between 1990 and 2003 is shown.  
    The horizontal bars above the data points indicate upper-limits.
The vertical line below ``HST" indicates the epoch of the HST observation described in \S~3.}
\end{center} 
%-----------------------------Figure End---------------------------------- 

\section{Short-term variability}

Transient black-hole binaries and transient Be/X-ray binaries both show short term (ms to ks) variability.
In black-hole systems the strength of aperiodic variability and the shape of the power-spectra are 
strongly dependent on the spectral state (McClintock \& Remillard 2003). During the giant outbursts of
Be/X-ray pulsars the broad band aperiodic variability is often stronger than the pulsations, with 
strong low-frequency quasi-periodic oscillations often appearing in the power spectra (Finger 2004). 
It is therefore of interest to quantify the short term aperiodic variability of \rxh ~~and to 
search for coherent pulsations.

We constructed short-term light curves of \rxh\ spanning individual observations and using the 
same extraction regions as described above to look for variability on short timescales.  
 For example, the light curve during the 1999 December 5 \cxo/ACIS-I observation is shown in Figure~4.
A total of 2264 counts were obtained in the 0.2--10.0 keV energy range, which were binned into 73 intervals of 500~s width. The light curve was steady during this observation according to a 
$\chi^2$ test against the constant count rate hypothesis (reduced $\chi^2$$=$0.995).
Figure~5 displays the \cxo/HRC light curve on 1999 September 10. 
Again using 500~s bins, the unweighted average is 1.4 counts per bin.
A $\sim$1500~s flare occurred at the beginning of the observation, lasted through three consecutive bins,
and contained a total of thirteen photons. 
A Monte Carlo simulation of a constant source with Poisson noise 
showed a probability of 6.0$\times 10^{-3}$ for two successive bins in the light curve with a total of 
more than ten photons making it unlikely that this flickering is a statistical fluctuation.
The light curves during the other observations are similar to these examples: 
A steady flux when \rxh\ is at its brightest and flickering during fainter episodes.

A power-spectrum analysis was conducted using event data from the 1999 December 5 \cxo/ACIS-I observation,
but no excess power above the Poisson noise was detected. The upper limit ($2\sigma$) to a 
constant power spectra in the $3 \times 10^{-5}-0.154$~Hz frequency range  is 
0.5 (rms/mean)$^2$~Hz$^{-1}$ at $10^{-3}$~Hz.  The upper limit for a zero-frequency Lorentzian with 
half-power cutoff at $10^{-2}$~Hz is 2.6 (rms/mean)$^2$~Hz$^{-1}$ at $10^{-3}$~Hz.

%-----------------------------Figure Start--------------------------------

\begin{center}
\includegraphics[angle=-90,width=\columnwidth]{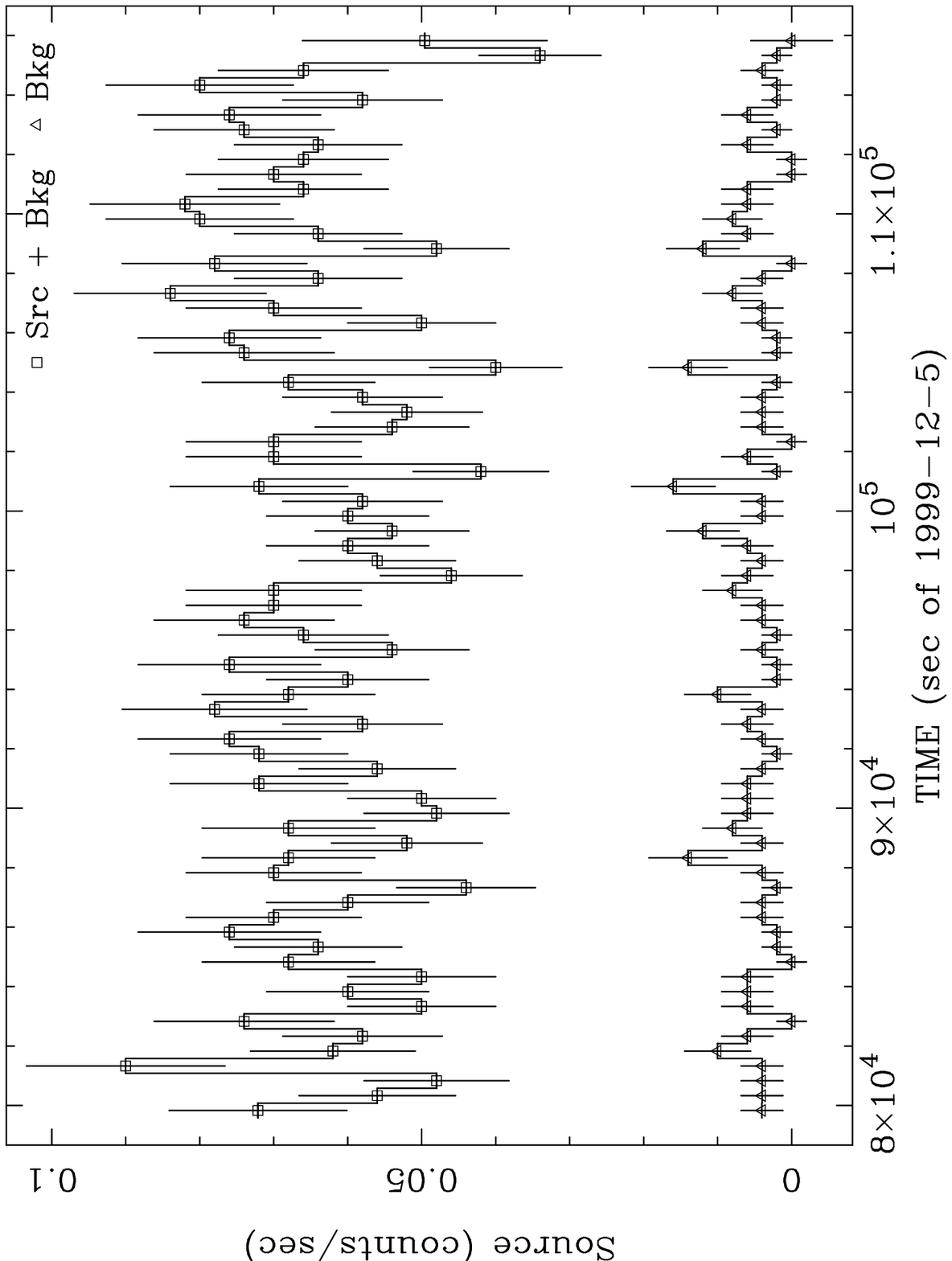}
\vspace*{-5cm}
\figcaption{The 0.2$-$10.0 keV  light curve of \rxh\ during the \cxo/ACIS-I 
    observation of  1999 December 05. 
    The source was bright and steady.
    Counts were binned into 500~s intervals.}
\end{center} 
%-----------------------------Figure End----------------------------------  

%-----------------------------Figure Start--------------------------------

\begin{center}
\includegraphics[angle=-90,width=\columnwidth]{Fig5.ps}
\figcaption{The full-band light curve of \rxh\ during the \cxo/HRC-I  
    observation of 1999 September 10. 
    The average source flux is about 10\% of the average on 1999 December 05 ({\sl cf.} Figure~4).  
    Counts were binned into 500~s intervals.}
\end{center}    
%-----------------------------Figure End---------------------------------- 

A search for coherent pulsations in the 3$\times$$10^{-5}$ to 0.154~Hz frequency range was made 
in the 1999 December 5 \cxo/ACIS-I data, but none were detected. The search used the $Z^2_n$ 
statistics (Buccheri et al.\ 1983) with both 
$n=1$ and $n=2$ Fourier terms.  The arrival times used were referred to the solar-system  
barycenter using the {\tt ciao} tool {\tt axbary } with the DE200 planetary  
ephemeris.
We obtained a 90\% confidence upper limit of 12\% on the 
rms pulse fraction.  The maximum values of $Z^2_n$ are consistent with Poisson statistics.  
(We note that this applies to simple pulse 
profiles limited to at most two Fourier terms.) 

A search for coherent pulsations with frequency up to 20 Hz was made in the 
1999 December 21 Chandra HRC-I data. Figure~6 shows the $Z^2_2$ statistic 
calculated from the 481  barycenter--corrected events extracted from a 6\arcsec\ by 3\arcsec\ ellipse centered 
on the source. 

The peak value was 47.39 at a barycentric frequency of 13.264391(3)~Hz.
To estimate the significance,
we used a Monte~Carlo algorithm to generate 33800 simulated data sets of 481 events each.
Forty-seven of these data sets had peaks exceeding 47.39 in the 0$-$20~Hz range 
indicating the probability of seeing such a peak in any one spectrum is $(1.4\pm0.2)\times10^{-3}$.
To test for instrumental effects, we also tested  635 background regions extracted from the HRC-I data, each containing 481
events, and found two with peaks above 47.39  for a probability of ($3.2\pm2.2)\times 10^{-3}$.

In Figure 7 we 
show the epoch-folded pulse profile, which has an rms pulse fraction of 31\% (two Fourier terms). 
The amplitude of the pulsations is only marginally increased by including a frequency rate term in the 
analysis. We measure a frequency rate of $(-2.4\pm1.6)\times 10^{-9}$~Hz\,s$^{-1}$.

We were unable to confirm these pulsations using the \ros/HRI data from 1995 July.   These data, which 
consists of 820 events (37 being background), are spread non-uniformly over an interval of 10 days. 
These data are therefore very sensitive to small changes in frequency  and therefore require a search in both frequency and  frequency-rate with a
large number of trial parameters. We used the $Z^2_2$ statistic to
search for pulsations at a frequency within 55 mHz of the Chanda HRC-I detection, and frequency rate
in the range of $-5.5\times 10^{-9}$ to $5.5\times 10^{-9}$ ~Hz\,s$^{-1}$.  The arrival times used were referred to the barycenter using the  
ROSAT {\tt FTOOLS} {\tt bct} and {\tt abc}.
We find an upper limit of 27\%
for any pulsations in this parameter range. If pulsations were present with the same pulse fraction as
seen in the Chandra data, with pulse phasing that could be accurately described by a frequency and frequency
rate in the search range, then it would have been clearly detected. The lack of detection could be 
due to a decrease of pulse fraction with decreasing energy, or the narrowness of the ranges of frequency and 
frequency-rate which was covered, which were limited by the amount of computation 
involved ($2\time 10^{10}$ evaluations of $Z^2_2$). In particular we point out that we failed to cover the 
frequency rate of $-1.0\times 10^{-7}$~~Hz\,s$^{-1}$ observed for \abe, and that the occurrence of a strong
orbital modulation (i.e. periastron passage) within the data interval would have prevented detection.

%-----------------------------Figure Start--------------------------------

\begin{center}
\includegraphics[angle=-90,width=\columnwidth]{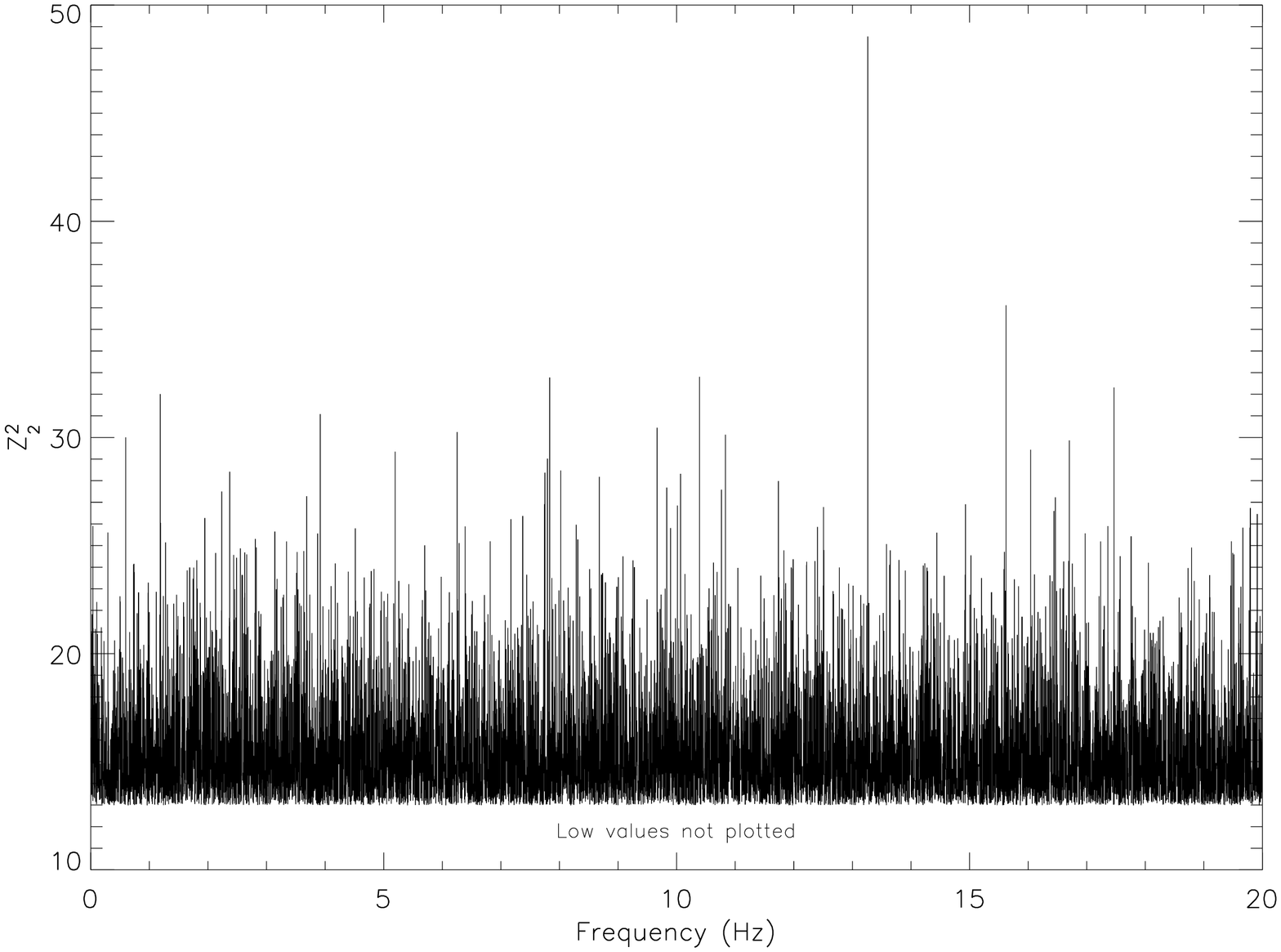}
\figcaption{
The $Z^{2}_{2}$ pulse search statistic for the Chandra/HRC-I data of 1999 December 12
showing a significant peak at 13.264391 Hz.}

\end{center} 
%-----------------------------Figure End---------------------------------- 

%-----------------------------Figure Start--------------------------------

\begin{center}
\includegraphics[angle=0,width=\columnwidth]{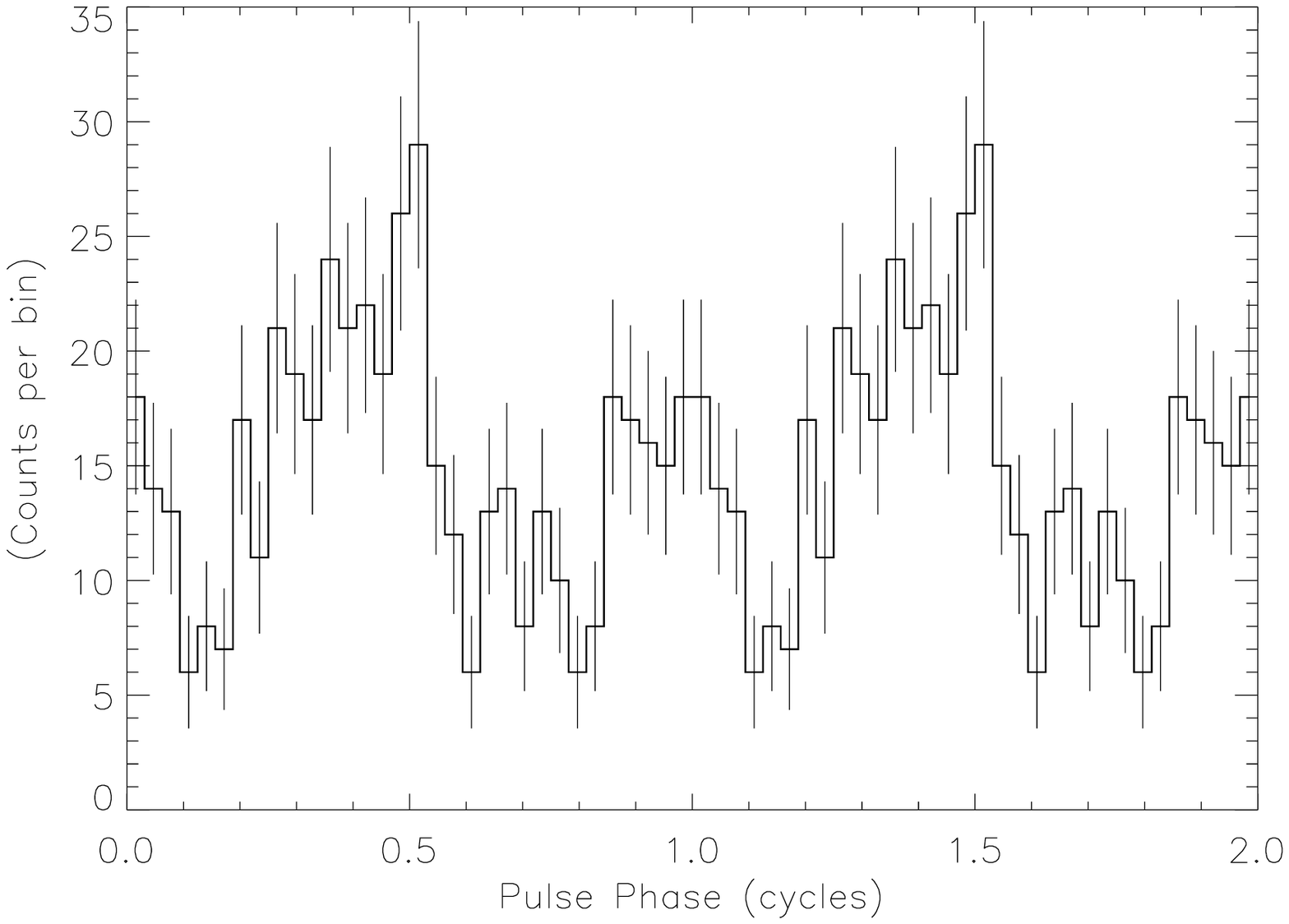}
\figcaption{Epoch-folded pulse profile of the Chandra/HRC-I lightcurve of \rxh\
obtained on 1999 December 12.}
\end{center} 
%-----------------------------Figure End---------------------------------- 

\section{An Optical Counterpart Candidate} 

There are two \hst\ Space Telescope observations with \rxh\ in the field of view.
These observations were taken in 1997 July 27
(MJD~=~50656; 195 days before the nearest X-ray observation on 1998 Feb 7, see Table~1).
The observations used the 
Wide Field Planetary Camera 2 (WFPC2) with the F555W and F814W filters, respectively.
Both consisted of three 60~s exposures that were combined to remove cosmic ray tracks. 

We first located the globular cluster HGHH-06 (Harris et al. \ 1992; Peng et al.\ 2004) in the HST image,
which gives an approximate correction to the HST coordinates.
With this correction, the Chandra source CXOU~J132520.1-430310 (Kraft et al. 2001)
lies very close to the bright object near pixel location (443, 594) shown
in Fig. 8.
If we assume that this source is the optical counterpart to CXOU~J132520.1-430310, we can then
 derive a relative HST/Chandra coordinate offset.
The offset in R.A. and Dec. of the catalogued position
of CXOU~J132520.1-430310 from its position in the \hst\ data is 
 $\Delta \alpha = -0.22$~s  and
 $\Delta \delta = +1.7\arcsec$.
We applied this offset to register the \hst\ and \cxo\ images. 

Figure~8 shows a 20\arcsec$\times$20\arcsec\ region of the \hst\ F555W image 
    centered at the expected position of \rxh\ with the 90\% confidence level \cxo\ error circle marked. 
There is an object within this error circle.   
Its observed magnitudes are  $24.1\pm 0.6$ and $23.1\pm 0.6$ in the F555W and F814W images, 
    respectively. 
The probabilities of finding a source of similar brightness in these two images 
due to statistical fluctuation alone are 1\% and 7\%, respectively.
By fitting a 2D circular Gaussian to both the F555W and F814W data
we find the position differs, at the 90 percent confidence level, by
1.5 pixel ($\sim$0.15 arcsec).
This relative offset could be due to the blending of two objects
in the F814W image.
 
Correcting for Galactic extinction yields an 
    absolute magnitude $M_{\rm F555W} =-4.5\pm 0.6$~mag
    and a color (F555W--F814W)$<$0.7$\pm$0.8~mag.
The absolute magnitude is consistent with that of a late O or early B main-sequence star.  
The extinction-corrected color 
 is redder than that of main-sequence O and B stars ({\sl cf.} V--I$=-0.4$ for a B0 star). 
The observed color could be made consistent with that of an OB star if $\sim$$2/3$ of the flux
measured in the F814W image is due to an unrelated second source.

%-----------------------------Figure Start--------------------------------

\begin{center}
\includegraphics[angle=-90,width=\columnwidth]{Fig8.ps}
\figcaption{
A 20\arcsec$\times$20\arcsec\ portion of the \hst/F555W image of Cen~A
 centered on the position of \rxh.
The (true) 0.48\arcsec \cxo\ error circle is shown at the astrometrically-corrected position of 
the transient. North and East are indicated at upper left.
}
\end{center} 
%-----------------------------Figure End----------------------------------  

\section{The nature of \rxh}

The inferred isotropic  luminosity of 1RXH~J132519.8 ~$-$430312 peaked above $10^{39}$~\ergl\ during two episodes
 separated by a 4.4 year interval within a 13 year observing span.
This places \rxh\  at the low end of the
 distribution of luminosities observed among the class of ultra-luminous X-ray sources
 (Fabbiano 1989; Swartz et al.\ 2004)
 and in the realm of luminous outbursts observed from several well-studied 
 Galactic and Magellanic Cloud X-ray sources (Liu, van Paradijs \& van den Heuvel~ 2000,  2001); primarily the (high-mass) Be/X-ray binaries 
 and the (usually low-mass) SXTs.

Many of the characteristics of 1RXH~J132519.8 ~$-$430312  presented here can be explained by comparison to either
of these classes of Galactic systems:
\rxh\ varied in flux by at least a factor of 1000 consistent with the range observed 
for both Be/X-ray binaries and SXTs (which can be $\sim$$10^6$ corresponding to 
quiescent flux levels far below detectability of \rxh).
The spacing and duration of outbursts of \rxh\ are not well constrained because of the sparsely 
scheduled observations, but the interval of 4.4 years between observed outbursts and the
$\sim$2 week outburst durations
are within the range of timescales observed in the Galactic Be/X-ray binaries and SXTs.
The peak luminosity of \rxh\ 
is in the range observed for black hole SXTs and within a factor
of a few of the most luminous Be/X-ray binary \abe.
The spectrum during the peak luminosity episode of \rxh\ is 
represented equally well by either a disk blackbody model like the high/soft or very high/soft state spectra of SXTs or  a cut-off powerlaw  of accreting pulsars\footnote{ 
Low cut-off energies have also been observed in the Be/X-ray binaries BSD 24--491 (Reig \& Roche 1999);  X Per (Schlegel et al. 1993); LSI +61235 (Haberl, Angelini \& Motch 1998).}.
Note that, due to the low number of photons detected,
the observations of \rxh\
cannot be used to detect the more moderate aperiodic behavior in the $10^{-2}$ Hz and higher
range typically seen in SXTs and in giant outbursts of Be/X-ray binaries although 
violent variability at low frequencies can be detected.

Two characteristics of \rxh\ can best be explained only in analogy to the Be/X-ray binaries.
The detection of a potential optical 
companion to \rxh\ with the brightness of a late O or early B star is just what is expected 
for a Be/X-ray binary. For example, the companion of \abe\ is a B2 III-IV star 
 (Charles et al. 1983) and would appear to be
23.9 mag if it was in Cen~A.
More importantly,
the coherent pulsations we have detected in \rxh\ can only be explained 
in conjunction with a spinning magnetized neutron star. 
In fact, the period of 75.4~ms is
remarkably similar to the 69.2~ms period of \abe\ (Skinner et~al. 1982). 
 This identification could be made more definitive if 
optical variability (preferably correlated with changes in X-ray flux) could be
established and if the X-ray pulsations could be confirmed with another outburst
observation (preferably with \xmm\ to also obtain another X-ray spectrum).

In many aspects, we cannot consider 1RXH~J132519.8 $-$430312 as 
    a perfect analog to A 0536$-$66.  This
source reached a luminosity of $8.5\times 10^{38}$~ergs~s$^{-1}$ for only 
a few hours, during an outburst that lasted a few days. A later 
super-Eddington outburst,
lasted for 2 weeks, but showed an $\sim$4 day e-folding decay (Skinner 
et al. 1980).
In contrast, \rxh\ ~was surprisingly constant for 10 days in the ROSAT 
outburst.
However, A~0538-66 does   provide an example of a neutron star radiating 
far above the Eddington luminosity.
Indeed, at the base of the accretion column the accretion rate per unit 
area must
be super-Eddington by a factor approching two orders of magnitude.

It is the magnetic field of the pulsar which produces an efficient segregation of 
mass inflow and photon outflow that allows such super-Eddington conditions to exist.
The observations can be used to place constraints on the neutron star's magnetic field.
The magnetic field must be strong enough to channel the accretion flow near the 
stellar surface. 
For a 1.4~\msun\ neutron star of radius $10^6$~cm, 
a polar strength greater than 6$\times$$10^{8}$~${\rm G}$ for a dipole magnetic field is needed to 
maintain the magnetosphere above
the neutron star surface when the accretion rate is sufficient to 
produce a luminosity of 10$^{39}$~ergs~s$^{-1}$.
%The observations also provide an upper limit to the magnetic field of the neutron star.
At low accretion rates, the %radius of a neutron star's 
magnetosphere can expand beyond the
radius at which a Keplerian orbit co-rotates with the neutron star resulting in material
embedded in the field being flung off by centrifugal forces. In this ``propeller regime,'' 
accretion to the surface is inhibited (Stella, White \& Rosner 1983).  Reduced levels of 
X-ray emission may occur due either to leakage through the 
lower centrifugal barrier near the spin axis, or due to accretion onto the magnetosphere 
(King \& Cominsky 1994). The transition to this regime occurs at a luminosity 

\[L_{\rm transition} = 4\times 10^{37} M_{1.4}^{-2/3} R_6^5 B_{12}^2 
\nu_{Hz}^{7/3}~~{\rm ergs~s}^{-1}\]

\noindent where  $M_{1.4}$ is the mass of the neutron star in units of 1.4~\msun, $R_6$ its radius in
units of $10^6$~cm, $B_{12}$ its polar field in units of $10^{12}$~G, and $\nu_{Hz}$ its spin frequency
in Hz. The flaring during the
1999 Sept 10 HRC-I observations could be explained by the unstable accretion expected during the transition
to the propeller regime (Spruit \& Taam, 1993). This would imply $B \approx 6\times 10^{10}$~G
for a 1.4~\msun\ neutron star and $10^6$~cm radius.

\rxh\ is classified as an ultraluminous X-ray source by virtue of its peak 
luminosity. How does \rxh\ compare to the general class of ULXs? During
its peak brightness episodes, the X-ray properties of \rxh\ are within the broad 
range of values observed from other ULXs. For example, 30\% of the ULXs 
surveyed by Swartz et al. (2004) have power law indices between 2.0 and 3.0
(although the average of the whole sample is only 1.7) and 86\% showed no significant variability
during times of observation. However, the fraction of ULXs that are 
{\sl transient} has not yet been documented so we do not know how common
this characteristic of \rxh\ is among the general class of ULXs. There have also
been several reports of late O or early B star companions to ULXs
(e.g. Liu et al. 2005) like the candidate counterpart to \rxh.

How common could Be/X-ray binaries like \rxh\ be within the ULX population?
A large number of Be-type binary
systems is expected in typical galaxies with modest star formation rates.
For example, the population synthesis studies of Dalton \& Sarazin (1995) 
predict $>$20000 potential Be-type binary systems 
exist in our Galaxy based on the observed Lyman continuum flux. 
In fact, most of the 130  known high-mass X-ray binaries in the Galaxy 
and Magellanic Clouds are Be/X-ray binaries (Liu et al. 2000). 
In addition, there  is a strong correlation between ULXs and recent star 
   formation in late-type galaxies. Be/X-ray binaries are short lived 
   and they are expected to be found in young stellar environments  
   such as those in late-type and interacting galaxies. 
 Thus, the potential certainly exists for Be/X-ray binaries to make a significant
 contribution to the ULX population, at least at the low luminosity end, and the ULX \rxh\ may be the first such example.

\begin{acknowledgements}
Our sincere thanks to the anonymous referee for valuable comments that helped to improve the 
paper.
This research has made use of the NASA/IPAC Extragalactic Database (NED) which
 is operated by the Jet Propulsion Laboratory, California Institute of
 Technology, under contract with NASA;
of data products from the Two Micron All Sky Survey, which is a joint project
 of the University of Massachusetts and the Infrared Processing and Analysis
 Center, funded by NASA and the NSF;
from the Multimission Archive (MAST) at the STScI operated by AURA under NASA
 contract NAS5-26555;
and from the Chandra Data Archive, part of the Chandra X-Ray Observatory
 Science Center (CXC) which is operated for NASA by SAO. 
Support for this research was provided in part by NASA under
 Grant NNG04GC86G issued through the Office of Space Science.

\end{acknowledgements}

\newpage

\onecolumn 

\vspace*{1cm}

\begin{center}
\small{
%\begin{tabular}
\begin{tabular}{cccccc}
\multicolumn{6}{c}{{\sc Table 1}} \\
\multicolumn{6}{c}{ Log of ROSAT, Chandra and XMM--Newton X-ray observations of Cen A (NGC~5128)} \\
\hline \hline
 \multicolumn{1}{c}{Date of obs.} & \multicolumn{1}{c}{MJD}& \multicolumn{1}{c}{ObsID} & \multicolumn{1}{c}{Satellite/Instrument} & \multicolumn{1}{c}{Exposure} & \multicolumn{1}{c}{0.1--2.4 keV Flux}\\
 \multicolumn{1}{c}{(yy--mm--dd)} &\multicolumn{1}{c}{(days)} && & (ksec)& \multicolumn{1}{c}{($10^{-15}$~\ergcms)} \\
\hline
90--07--27 & 48099 & 150004& ROSAT/HRI&19.52&$<$~10.0$^{a}$\\
92--01-26 & 48647 & 700012& ROSAT/PSPC&13.48& $<$~23.0$^{a}$\\
94--08--10& 49574 & 701571& ROSAT/HRI& 64.11& $<$~13.0$^{a}$\\
95--07-13&49911 & 701924& ROSAT/HRI&4.90&1280$\pm$100\\
95--07--14&49912 &701925 & ROSAT/HRI&5.33& 1460$\pm$110\\
95--07--18&49916 &701926 & ROSAT/HRI&5.72&1310$\pm$100\\
95--07--22&49920 & 701927& ROSAT/HRI& 4.09&970$\pm$100\\
95--07--23&49921 & 701928& ROSAT/HRI& 4.06&1310$\pm$110\\
98--02--07&50851 & 704206& ROSAT/HRI& 17.22&$<$~27.0$^{a}$\\
99--09--10&51431 & 463& Chandra/HRC-I&19.51& 21.4$\pm$5.3\\
99--09--10&51431 & 1253& Chandra/HRC-I&6.81& 14.3$\pm$8.5\\
99--12--05&51517 & 316& Chandra/ACIS-I& 35.72 &303.1$\pm$6.0\\
99--12--21&51533 & 1412& Chandra/HRC-I&14.97& 316.0$\pm$14.0\\
00--05--17&51681 & 962& Chandra/ACIS-I &36.50&$<$~1.7$^{a}$\\
01--02--02&51942&93650201&XMM--Newton/MOS&23.02&$<$~3.8$^{a}$\\
01--05--09&52038 & 1600& Chandra/ACIS-S/HETG&46.84& $<$~2.1$^{a}$\\
01--05--21& 52050 &1601& Chandra/ACIS-S/HETG&51.50&3.5$\pm$1.9\\
02--02--06&52311&93650301&XMM--Newton/MOS/PN&8.89&$<$~3.8$^{a}$\\
02--09--03& 52520& 2978& Chandra/ACIS-S& 44.59&$<$~1.3$^{a}$\\
03--09--14& 52896& 3965& Chandra/ACIS-S& 49.52&$<$~1.3$^{a}$\\
\hline
\multicolumn{6}{l}{$^a$The upper limit is at the 2$\sigma$ level. }\\
\end{tabular}
} %end \small
\end{center}

\bigskip

\begin{center}
\small{
\begin{tabular}{lcccccc}
\multicolumn{7}{c}{{\sc Table 2}} \\
\multicolumn{7}{c}{ X-ray spectral parameters of  1RXH J132519.8$-$430312 
      on 1999 December 5   } \\
\hline \hline
 \multicolumn{1}{c}{XSPEC} & \multicolumn{1}{c}{$N_{\rm H}$} & \multicolumn{1}{c}{$\Gamma$} & 
\multicolumn{1}{c}{$E_{cut}$} &\multicolumn{1}{c}{$T_{\rm  e, in}$} & \multicolumn{1}{c}{$L_{\rm X}^a$} & $\chi^2$/dof \\
 \multicolumn{1}{c}{Model} & \multicolumn{1}{c}{($10^{20}$ cm$^{-2}$)} & &\multicolumn{1}{c} {(keV)} &\multicolumn{1}{c}{(keV)} & \multicolumn{1}{c}{($10^{39}$~ergs~s$^{-1}$)} \\
\hline
Power-law         & $35.0^{+5.0}_{-5.0}$  & $2.63^{+0.13}_{-0.12}$&---  & --- & $1.61\pm0.11$  & 112.3/82\\
Bremsstrahlung    & $18.9^{+3.4}_{-3.2}$  & ---  &---&$2.33^{+0.25}_{-0.22}$  & $1.20\pm0.10$  &  80.1/82 \\
Raymond-Smith$^b$ & $18.9^{+3.3}_{-3.0}$  & ---  &---&$2.32^{+0.24}_{-0.21}$  & $1.17\pm0.09$  &  80.1/81 \\
Mekal$^b$         & $19.0^{+3.7}_{-2.7}$  & ---  &---&$2.34^{+0.21}_{-0.23}$  & $1.18\pm0.09$  &  79.9/81 \\
Disk blackbody    & 8.63 & ---&---   &$0.93^{+0.04}_{-0.04}$  & $1.09\pm0.08$  & 71.8/83 \\
Cut-off powerlaw$^c$ &$21.7^{+6.7}_{-3.2}$&$2.01^{+0.22}_{-0.27}$  &3.31$\pm$0.35&$1.87^{+0.73}_{-0.40}$$^d$& $1.14\pm0.16$&
70.2/80\\
\hline
\multicolumn{7}{l}{$^a$ Intrinsic isotropic luminosity in the 0.5$-$8.0 keV band.}\\
\multicolumn{7}{l}{$^b$Elemental abundance is a free parameter which converges to $Z$$=$0.}\\
\multicolumn{7}{l}{$^c$ Powerlaw~$*$~highecut model in XSPEC}\\
\multicolumn{7}{l}{$^d$ Folding energy in keV.}\\
\end{tabular}
} %end \small
\end{center}

%%%%%%%%%%%%%%%%%%%%
\end{document}